# Benchmarking triple stores with biological data

Vladimir Mironov[1], Nirmala Seethappan[1,2], Ward Blondé[3], Erick Antezana[1], Bjørn Lindi[2], and Martin Kuiper[1]

[1] Dept. Biology, Norwegian University for Science and Technology (NTNU), Trondheim, Norway
{Martin.Kuiper, Erick.Antezana Vladimir.Mironov}@bio.ntnu.no
[2] High Performance Computing, Norwegian University for Science and Technology (NTNU), Trondheim, Norway
Bjorn.Lindi@ntnu.no, s.nirmala@gmail.com
[3] Dept. Applied Mathematics, Biometrics and Process Control, Ghent University, Ghent, Belgium
Ward.Blonde@ugent.be

**Abstract.** We have compared the performance of five non-commercial triple stores, Virtuoso-open source, Jena SDB, Jena TDB, SWIFT-OWLIM and 4Store. We examined three performance aspects: the query execution time, scalability and run-to-run reproducibility. The queries we chose addressed different ontological or biological topics, and we obtained evidence that individual store performance was quite query specific. We identified three groups of queries displaying similar behavior across the different stores: 1) relatively short response time, 2) moderate response time and 3) relatively long response time. OWLIM proved to be a winner in the first group, 4Store in the second and Virtuoso in the third. Our benchmarking showed Virtuoso to be a very balanced performer – its response time was better than average for all the 24 queries; it showed a very good scalability and a reasonable run-to-run reproducibility.

## 1 Introduction

Semantic Web technologies are increasingly being adopted by the scientific community, and Life Sciences researchers are no exception [1]. Our perspective is from the Life Sciences, and we have previously built two semantically integrated knowledge bases [2,3]. Semantic Web technologies open a new dimension to data integration, with various solutions, such as format standardization at the source (ontologies and uniform semantics), a sound scalability system, and an advanced exploratory analysis (e.g. automated reasoning), to overcome some of the current limitations. An increasing number of principal biological data providers, such as

UniProt [4], have started to make their data available in the form of triples (commonly represented in the Resources Description Framework (RDF) language [5]). Access to data in RDF format typically is facilitated via so-called endpoints. Those endpoints allow querying by SPARQL [6], the standard query language that allows users experienced in this query language to fetch information from resources holding RDF triple stores – a collection of terms and their interrelationships.

### 1.1 Triple stores

Currently, there are several triple store solutions [7] to store information represented in RDF format. Although most of them are not targeted towards a specific domain, some of them have been readily adopted by biological data handlers who expected to find in them a means to overcome some of the limitations of classical storage solutions (mainly based on relational database management systems).

The development of triple stores has flourished during the last 5 years. Currently, there are more than 20 systems available [8]. Both the academic and private sectors have been involved in developing these triple stores. This race has created a healthy competition to excel in querying and loading performance, scalability, and stability. In particular, the semantic web community has been also challenging the usage of triple stores by promoting open contests and demonstrating Semantic Web applications [9]. It is encouraging for the scientific community that many of these triple stores are freely available for academic use.

### 1.2 Benchmarking efforts

Much of the benchmarking done previously on triple stores was based on artificial data or a set of triples that could at best only mimic a realistic ontology. Among the "standard" sets used are: the Lehigh University Benchmark (LUBM [10]) and the Berlin SPARQL Benchmark (BSBM, [11]). Other studies, such as the one performed by UniProt [4], demonstrated the current limitations of some triple stores [12].

Here, we present "the NTNU benchmark", which is the work we undertook using five popular triple store implementations, and report the outcome of this benchmarking. In comparison with previous benchmarkings [13], we used two additional stores not included previously (Swift OWLIM and 4Store) and instead of (artificial) computationally generated data, we used biologically relevant real life data from our Cell Cycle Ontology knowledge base [2].

## 2  Benchmarking

### 2.1  Software

The set of triple store implementations included Virtuoso OpenSource 6.0.0, Swift OWLIM 2.9.1, 4Store 1.0.2, Jena SDB 1.3.1, Jena TDB 0.8.2. The stores were run under Centos 5 operating system. The details of software configuration are available on request.

### 2.2  Hardware

The benchmarking was performed on a Dell R900 machine with 24 Intel(R) Xeon(R) CPUs (2.66GHz). The machine was equipped with 132G main memory and 14x500GB 15K SAS hard drives.

### 2.3  Querying

The ten graphs constituting the Cell Cycle Ontology (CCO) [2], in size ranging from 356903 to 3170556 triples, were used for benchmarking. The graphs were queried with 24 SPARQL queries from the library of queries on the CCO web site (http://www.semantic-systems-biology.org/cco/queryingcco/sparql). The queries were executed on each of the graphs sequentially from query Q1 through Q24. The experiments were replicated three times. Prior to each experiment the contents of the stores were completely cleared and uploaded anew. The average response time and the corresponding relative standard errors (RSE) for these three observations were computed for all the data points (24 queries and 10 graphs, available as supplementary material) and used to aggregate the data for Tables 2 and 3 and Figures 1 and 2.

## 3  Results and Discussion

The most salient features of the queries used for our benchmarking are summarized in Table 1.

Table 1. Overview of the query features. The selected 24 queries (Q1 through Q24) were used to evaluate the triple stores' responsiveness with respect to various query features (e.g. REGEX). The table shows the full set of queries and the features used.

| | Simple Filters | More than 8 triple patterns | OPTIONAL operator | LIMIT modifier | ORDER BY modifier | DISTINCT modifier | REGEX operator | UNION operator | COUNT operator |
|---|---|---|---|---|---|---|---|---|---|
| Q1 | | | | | | x | | | |
| Q2 | | | | | | | | x | |
| Q3 | x | x | x | | | x | x | | |
| Q4 | | | | | | | | | |
| Q5 | | | | | | | | | |
| Q6 | | | | | | | | | |
| Q7 | x | | | | | | x | | |
| Q8 | | | | | | x | | | |
| Q9 | x | | | | | | | | |
| Q10 | x | x | | | | x | x | | |
| Q11 | | | | | | | | | |
| Q12 | | | | | | | | x | |
| Q13 | | x | x | | | x | | x | |
| Q14 | | | | | x | | | | |
| Q15 | | | | | | | | | |
| Q16 | | | | | | | | | |
| Q17 | | | | | | x | | | x |
| Q18 | x | x | | | x | | x | x | |
| Q19 | | | | | x | | | x | x |
| Q20 | | | | | x | | | | x |
| Q21 | | | | | x | | | | x |
| Q22 | | | | | | | | | |
| Q23 | | | | | | | | | |
| Q24 | x | | | x | | | x | | |

As can be seen from Table 1, this collection of queries encompasses a broad range of features and combinations thereof. This ensures a comprehensive assessment of the performance of the triple stores.

In order to get a bird's eye view on the performance of the stores we aggregated the response times into the single cumulative total response time and estimated the average relative standard errors for each of the stores (Table 2). (Please note that OWLIM does not support the COUNT operator, therefore the values for this store do not include data for queries Q17, Q19, Q20. Q21).

Table 2. Response times (in seconds) averaged over the three replicates and summed over the 24 queries and 10 graphs. RSE – the relative standard error for the three replicates averaged over all the data points (24 queries and 10 graphs).

| Store | Total time | RSE |
|---|---|---|
| Virtuoso | 203.939 | 0.053 |
| Jena SDB | 730.492 | 0.020 |
| Jena TDB | 1445.572 | 0.235 |
| OWLIM | 14257.964 | 0.156 |
| 4Store | 47566.530 | 0.097 |

The total execution time varied over a very broad range and some of the stores (most notably Jena TDB and OWLIM) displayed an unexpectedly high run-to-run variability. On the basis of these data Virtuoso emerges as an overall winner, with by far the best total execution time and a relatively small run-to-run variation. However, the picture changes radically when we look into the query- specific behavior (Table 3).

Table 3. Average response time in seconds summed over the 10 graphs and sorted by the average execution time. The slowest response is highlighted in red, the fastest in green. The queries are sorted in the order of the response time averaged over the 5 stores (Avg).

| Query | Virtuoso | Jena SDB | Jena TDB | 4Store | OWLIM | Avg |
|---|---|---|---|---|---|---|
| Q5 | 2.639 | 13.446 | 11.000 | 1.526 | 0.408 | 5.804 |
| Q23 | 5.630 | 13.343 | 10.454 | 1.343 | 0.009 | 6.156 |
| Q11 | 5.343 | 13.339 | 10.703 | 1.419 | 0.011 | 6.163 |
| Q16 | 5.617 | 13.345 | 10.825 | 1.346 | 0.009 | 6.228 |
| Q15 | 6.163 | 13.342 | 10.544 | 1.390 | 0.018 | 6.291 |
| Q22 | 5.170 | 13.709 | 10.981 | 1.428 | 0.173 | 6.292 |
| Q4 | 5.916 | 13.348 | 10.773 | 1.539 | 0.017 | 6.319 |
| Q8 | 7.094 | 13.336 | 10.449 | 1.577 | 0.049 | 6.501 |
| Q12 | 7.198 | 13.373 | 10.731 | 1.400 | 0.030 | 6.546 |
| Q2 | 7.281 | 13.337 | 10.768 | 1.438 | 0.052 | 6.575 |
| Q6 | 4.054 | 14.523 | 10.573 | 1.779 | 2.020 | 6.590 |
| Q19 | 2.065 | 13.390 | 9.795 | 1.326 | | 6.644 |
| Q9 | 5.820 | 13.711 | 10.699 | 2.133 | 1.067 | 6.686 |
| Q21 | 3.379 | 13.335 | 9.818 | 1.316 | | 6.962 |
| Q10 | 4.679 | 13.757 | 11.676 | 2.529 | 4.664 | 7.461 |
| Q17 | 5.648 | 13.390 | 10.119 | 1.350 | | 7.627 |
| Q20 | 6.110 | 13.387 | 10.686 | 1.315 | | 7.875 |
| Q1 | 1.897 | 18.064 | 14.258 | 1.647 | 8.024 | 8.778 |
| Q13 | 1.658 | 52.545 | 14.156 | 1.569 | 0.034 | 13.992 |
| Q24 | 2.813 | 24.719 | 38.619 | 14.366 | 27.242 | 21.552 |
| Q7 | 2.617 | 26.519 | 39.248 | 14.110 | 28.996 | 22.298 |
| Q14 | 5.775 | 13.338 | 46.433 | 1.401 | 91.894 | 31.768 |
| Q3 | 3.358 | 30.476 | 27.049 | 3.654 | 1121.702 | 237.248 |
| Q18 | 22.840 | 76.013 | 493.596 | 24999.569 | 8325.734 | 6783.550 |

The table makes clear that all the stores behave in a query specific manner. A highly query-specific behavior has been also observed by Bizer and Schultz [14]. However, a couple of common trends are visible. OWLIM is by far the best performer with the relatively short response time queries; 4Store shows the best performance with the moderate response time queries; and Virtuoso is doing best of all with the long response time queries. Jena SDB is consistently the slowest store with all the short and moderate response time queries. Additionally, it should be noted that for OWLIM and 4Store the cumulative values in the Table 2 are dominated by outliers – query Q18 for 4Store and queries Q14, Q3, Q18 for OWLIM. The only common feature of the queries Q14 and Q18 is the ORDER BY modifier, not used by any other queries. The list of features shared by the queries Q3 and Q18 includes simple filters, more than 8 triple patterns and a REGEX operator. At present it is not possible to determine which of these features or a combination thereof are responsible for the long execution time.

Finally, we wanted to compare the stores with respect to their scalability. The averaged response times were summed over all the queries (except for the queries Q17, Q19, Q20. Q21 for OWLIM) and plotted against the total number of triples in the graphs (Figure 1).

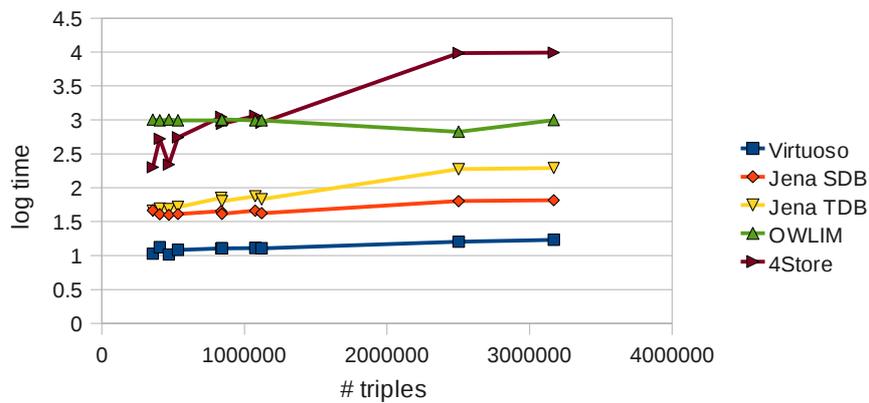

**Fig. 1.** Average response time in seconds summed over the 24 queries. The response times were averaged over the three replicates and summed over all the queries (except for the queries Q17, Q19, Q20. Q21 for OWLIM due to the lack of support for the COUNT operator) and plotted against the total number of triples in the graphs.

As can be seen from the figure OWLIM scales up extremely well, with Virtuoso and Jena SDB as second best. 4Store demonstrated the poorest performance with respect to scalability. However, as pointed out earlier, the behavior of OWLIM and 4Store is strongly affected by a few outliers. Therefore, to eliminate the impact of the outliers we excluded the three slowest queries Q3, Q14 and Q18 from the plot (Figure 2).

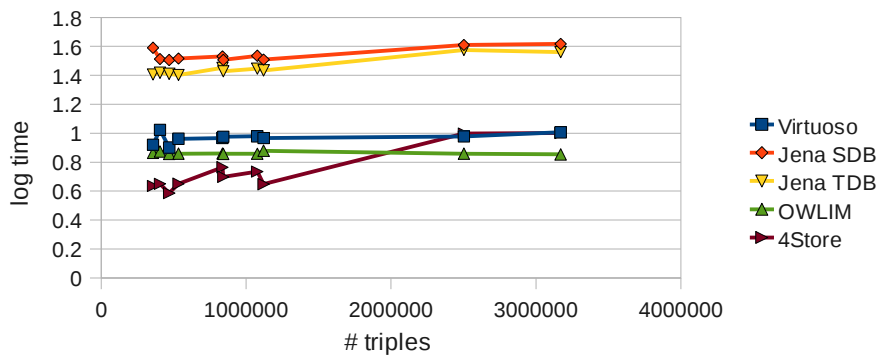

**Fig. 2.** Average response time in seconds summed over the 21 queries. The same as Fig. 1 but omitting the queries Q3, Q14 and Q18.

Although the mutual arrangement of the individual graphs on the plot changed in favor of OWLIM and 4Store, the conclusion drawn previously about the scalability did not change.

## 4    Conclusions

We have compared the performance of five popular triple stores, Virtuoso-open source, Jena SDB, Jena TDB, Swift OWLIM and 4Store, in three aspects – the query execution time, scalability and run-to-run reproducibility. According to our results there is no absolute winner within this set of stores. Instead, the performance seems to be quite query-specific. Nevertheless, it was possible to identify three groups of queries displaying similar behavior with respect to the different stores: 1) relatively short response time, 2) moderate response time and 3) relatively long response time. OWLIM proved to be a winner in the first group, 4store in the second and Virtuoso in the third. Virtuoso emerged from our benchmarking as a very balanced performer – its response time was better than average for all the 24 queries; it showed a very good scalability and a reasonable run-to-run reproducibility. Even though in our study we used only moderately large triple stores, others demonstrated that Virtuoso excels when confronted with much larger stores, up to 100-200 M triples [14,15]. We conclude that Virtuoso is well suited for managing large volumes of biological data. This conclusion is further corroborated by the successful deployment of Virtuoso in our BioGateway project [16] where it gracefully supports querying of ~1.8 billion triples.

**Data availability.** The rdf files used for uploading the triple stores are downloadable from the CCO web site [17].

**Acknowledgments.** The authors appreciate the assistance of E. Jensen with setting up the triple stores used in this benchmarking. V. Mironov was supported by FUGE Midt Norge.